\documentclass[prb,showpacs,amsmath,amssymb,twocolumn,letterpaper]{revtex4}
\usepackage{amssymb}
\usepackage{graphicx}
\usepackage{color}
\usepackage{dcolumn}
\usepackage{bm}

\begin{document}
\title{High-pressure, transport, and thermodynamic properties of CeTe$_3$}
\author{D.\ A.\ Zocco}
\affiliation{Department of Physics and Institute for Pure and
Applied Physical Sciences, University of California, San Diego, La
Jolla, California 92093, USA}
\author{J.\ J.\ Hamlin}
\affiliation{Department of Physics and Institute for Pure and
Applied Physical Sciences, University of California, San Diego, La
Jolla, California 92093, USA}
\author{T.\ A.\ Sayles}
\affiliation{Department of Physics and Institute for Pure and
Applied Physical Sciences, University of California, San Diego, La
Jolla, California 92093, USA}
\author{M.\ B.\ Maple}
\affiliation{Department of Physics and Institute for Pure and
Applied Physical Sciences, University of California, San Diego, La
Jolla, California 92093, USA}
\author{J.-H.\ Chu}
\affiliation{Department of Applied Physics, Geballe Laboratory for
Advanced Materials, Stanford University, Stanford, California 94305,
USA}
\author{I.\ R.\ Fisher}
\affiliation{Department of Applied Physics, Geballe Laboratory for
Advanced Materials, Stanford University, Stanford, California 94305,
USA}
\date{\today}

\begin{abstract}

We have performed high-pressure, electrical resistivity and
specific heat measurements on CeTe$_{3}$ single crystals. Two
magnetic phases with non-parallel magnetic easy axes were detected
in electrical resistivity and specific heat at low temperatures.
We also observed the emergence of an additional phase at high
pressures and low temperatures and a possible structural phase
transition detected at room temperature and at 45 kbar which can
possibly be related with the lowering of the charge density wave
transition temperature known for this compound.
\end{abstract}

\pacs{62.50.-p, 71.27.+a, 75.25.+z, 71.45.Lr}

\maketitle

\section{Introduction}

Charge- and spin-density waves (CDWs and SDWs) frequently occur in
low-dimensional materials, and are driven by electron-phonon and
electron-electron interactions. These phases are formed by nesting
of the Fermi surface, which can be perfect in one-dimensional
systems.\cite{gruner94} A perfect nesting CDW system refers to the
situation where all the electrons near the Fermi surface are excited
with the same $q$ vector of a particular phonon mode. On the other
hand, incomplete nesting takes place for higher dimensional
materials, for which a density wave gap opens only over certain
regions of the Fermi surface. Rare-earth tritellurides $R$Te$_{3}$
($R$ = La-Tm, except for Eu) constitute a class of
quasi-two-dimensional materials that has recently attracted a
considerable amount of attention because the electronic properties
of these materials can be changed by substituting one rare-earth
element for another, making them ideal candidates for investigating
the properties of the CDW state.\cite{dimasi95, brouet04, brouet08}
The rare-earth tritellurides crystallize in the NdTe$_{3}$ structure
that belongs to the space group $Cmcm$ (No. 63); the structure
consists of alternating double layers of nominally square-planar Te
sheets and corrugated double $R$Te layers and forms a weakly
orthorhombic lattice.\cite{norling66} In this standard space group
denomination, the $b$-axis is oriented perpendicular to the
$ac$-planes, and the average lattice parameters for all the
lanthanide series are (a, b, c) $\sim$ (4, 26, 4) \AA. It is evident
that these compounds are electronically anisotropic, with the Te
planes quite decoupled from the $R$Te slabs.
\cite{komoda04,pearsons97} For this family of materials, the lattice
modulation is characterized by a single in-plane wave vector, which
has approximately the same value for all the rare earths (2c*/7,
with c*=2$\pi$/c). \cite{dimasi95}

It has been shown that the application of chemical pressure
reduces the CDW ordering temperature from values above 450 K for
(La,Ce,Nd,Pr)Te$_{3}$ to 244 K for TmTe$_{3}$.\cite{sacchetti06,
ru08,ru08_2} Moreover, a reduction with chemical pressure of the
single particle excitation frequency characteristic of the CDW
state is accompanied by a decrease in the fraction of the Fermi
surface that remains gapped, driving the samples towards a state
of enhanced metallicity. This behavior was also observed in
CeTe$_{3}$ with the application of external pressure,
\cite{sacchetti07, lavagnini09} extending the study of the above
mentioned phenomenon to even smaller lattice parameters than
attainable through chemical pressure.

A second CDW ordering temperature has recently been discovered for
the compounds with smaller lattice parameters (Tm, Er, Ho,
Dy).\cite{ru08} In this case, the CDW is characterized by a
wavevector transverse to the first one, and of larger value
(a*/3). This phase occurs at lower temperatures, dropping below 50
K for DyTe$_{3}$, and it increases with the application of
chemical pressure. The shift of our attention to CDW formation at
low temperatures allows us to consider the effects caused on this
state by other competing types of order. Cerium-based compounds
frequently display an enhancement of their electronic effective
mass at low temperatures caused by the strong hybridization of the
localized 4\textit{f} and conduction electron states and produce a
variety of ground states, such as localized moment magnetic
order\cite{sullow99} and superconductivity,\cite{jung03} with many
of these phases induced at high pressures.\cite{mathur98} The
competing interaction of the CDW with some of these strongly
correlated electron states by tuning chemical composition,
pressure or magnetic field is of particular interest in these
materials.\cite{berthier76} In this report, we present
high-pressure electrical transport measurements on CeTe$_{3}$,
along with the results of a sub-kelvin specific heat experiment at
ambient pressure and high magnetic fields. We have found that two
magnetic phases occur below 20 K, with non-parallel magnetic easy
axes, as can be inferred from the additional transport
measurements made in fields for different angles. A possible
structural phase transition suggested by features in the
electrical resistivity at room temperature and at a pressure of 45
kbar, along with the low temperature features detected at high
pressures, may indicate the reduction of the CDW transition
temperature below 300 K for the range of pressures used in our
experiments.

\begin{figure}[tbp]
{\includegraphics[width=2.8in]{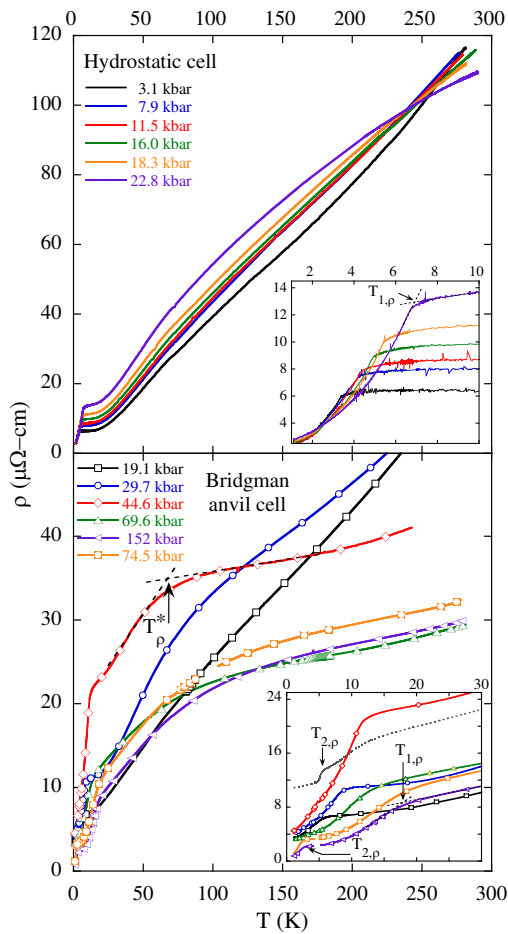}} \caption{(Color online)
Electrical resistivity versus temperature at various pressures for
CeTe$_{3}$ single crystals. \textit{Upper panel}: Hydrostatic cell
experimental results. The \textit{inset} shows the low temperature
range, displaying the onset of the magnetic order as T$_{1,\rho}$.
\textit{Lower panel}: Bridgman-anvil cell results. The
\textit{inset} shows T$_{1,\rho}$, along with the new ordering
temperature T$_{2,\rho}$, indicated in the figure for the first
Bridgman run (for the 152 kbar curve) and for the second Bridgman
run (in black, dashed curve).} \label{RHOvsTvsP}
\end{figure}

\section{Experimental details}

Single crystals of CeTe$_{3}$ were grown by slow cooling of a binary
melt as described elsewhere.\cite{ru06} Electrical resistivity
measurements under pressure were performed throughout the
temperature range 1.1 K $\leq$ T $\leq$ 300 K, employing two
different techniques. In the first technique, pressure was applied
with a beryllium-copper, piston-cylinder clamped cell using a Teflon
capsule filled with a 1:1 mixture of n-pentane:isoamyl alcohol as
the pressure transmitting medium to ensure hydrostatic conditions
during pressurization at room temperature. The pressure in the
sample chamber was inferred from the inductively determined,
pressure-dependent superconducting critical temperature of a lead
manometer,\cite{bireckoven88} and reached a maximum value of 23
kbar. In the second technique, pressure was applied in a
beryllium-copper Bridgman anvil clamped cell using solid steatite as
the quasi-hydrostatic pressure transmitting medium. The pressure was
determined from the superconducting transition of a strip of lead
foil placed adjacent to the sample and measured using a 4-lead
resistive method. With this technique, a maximum pressure of 152
kbar was attained in a first attempt and 124 kbar in a second run.
Pressure gradients were inferred from the width of the lead
superconducting transition. These gradients were as large as 2\% and
10\% of the total pressure for the piston-cylinder and the
Bridgman-anvil cell experiments, respectively. In both cases, the
electrical resistance in the $ac$-plane was measured using a 4-lead
technique and a Linear Research Inc. LR-700 AC resistance bridge.
Resistivity measurements at ambient pressure were obtained using a
Quantum Design Physical Property Measurement System (PPMS)
throughout the temperature range 1.9 K $\leq$ T $\leq$ 20 K and for
magnetic fields ranging from 0 to 9 T, applied both parallel and
perpendicular to the $b$-axis of the crystals.

The specific heat C of two single crystals with total mass of 7.5 mg
was measured as a function of temperature T from 0.65 to 200 K using
a ${}^{3}$He semiadiabatic calorimeter and a standard heat pulse
technique for magnetic fields up to 5 T applied along the $b$-axis.

\begin{figure}[tbp]
{\includegraphics[width=3.4in]{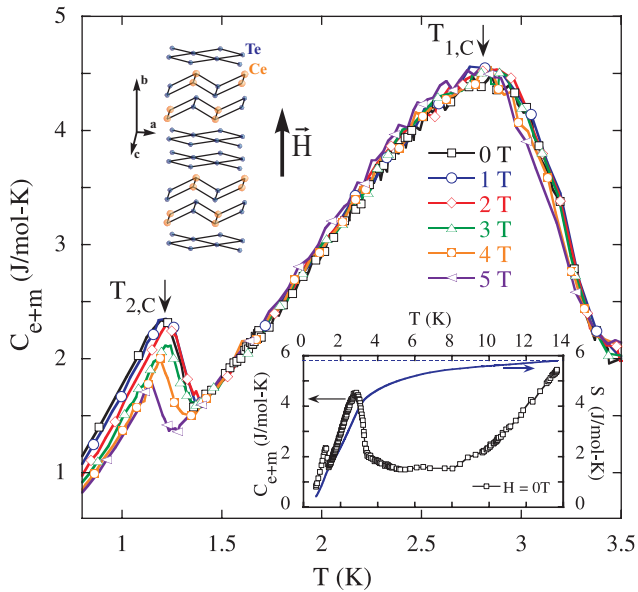}} \caption{(Color online)
Electronic and magnetic contribution to the specific heat for
different magnetic fields applied along the $b$-axis of the
CeTe$_{3}$ crystals. T$_{1,C}$ corresponds to the
field-independent ordering temperature centered near 3 K, while
T$_{2,C}$ denotes the field-dependent feature below 1.5 K. In the
lower \textit{inset}, the solid curve (blue) corresponds to the
electronic and magnetic entropy in zero magnetic field and the
horizontal dashed line corresponds to the value Rln2 J/mol-K. The
upper \textit{inset} shows a schematic diagram of the crystal
structure of CeTe$_{3}$ and the direction of the applied magnetic
field in the specific heat experiment.} \label{CvsT}
\end{figure}

\section{Results and Discussion}

Electrical resistivity measurements as a function of temperature
for different values of applied pressure are plotted in
Fig.~\ref{RHOvsTvsP}. The upper panel shows data obtained from the
hydrostatic cell experiment for pressures up to 23 kbar, while the
lower panel displays data taken in the Bridgman cell experiments
described in the previous section and for pressures up to 152
kbar. At low pressures (hydrostatic cell, upper panel), the sample
behaves as previously reported by Ru \textit{et al.},\cite{ru06}
although no appreciable local minimum at 10 K has been seen for
these initial values of pressure. In our measurements, the
resistivity decreases monotonically throughout the entire
temperature range, which is more evident for the highest pressures
obtained in the Bridgman anvil cell (lower panel). Below 100 K, a
broad hump denoted as T$^{*}_{\rho}$, which is clearly noticeable
for the higher pressures, moves to lower temperatures to a value
of 55 K at 50 kbar, after which it remains mostly unchanged for
the higher pressures. This feature occurring at T$^{*}_{\rho}$ is
suggestive of the appearance of the charge density wave order,
which is supported by recent x-ray diffraction data obtained for
CeTe$_{3}$ under pressure,\cite{sacchetti08} where it is clearly
seen that the onset of the CDW state occurs below room temperature
at 30 kbar. Nevertheless, the effects of the crystalline electric
field or the onset of Kondo coherence should not be ruled out,
taking into account the strong hybridization of the localized
4\textit{f} orbitals with the conduction band that usually takes
place in cerium-based compounds. A lower temperature feature,
labeled as T$_{1,\rho}$ was first reported by Iyeiri \textit{et
al.}\cite{iyeiri03} and later in the above mentioned work by Ru
\textit{et al.}. They attribute this feature to a transition to an
antiferromagnetic state, given the negative Curie-Weiss
temperatures obtained from magnetic susceptibility measurements.
We found that this ordering temperature increases from 3 to 13 K
under pressure, as can be seen in the \textit{insets} of the upper
and lower panels of Fig. \ref{RHOvsTvsP}. No appreciable change in
T$_{1,\rho}$ is observed for pressures above 50 kbar. For
pressures below 23 kbar, power-law fits to the resistivity curves
below T$_{1,\rho}$ yielded exponent values averaging 2.2 $\pm$
0.1.

\begin{figure}[tbp]
{\includegraphics[width=3.4in]{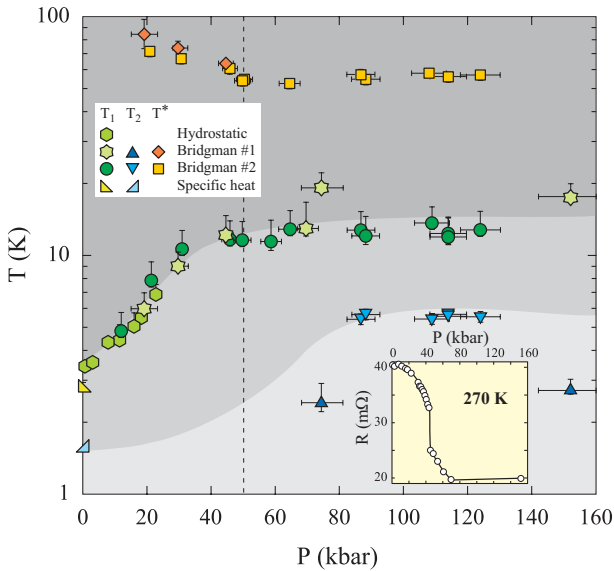}} \caption{(Color online)
Temperature versus pressure phase diagram for CeTe$_{3}$ (note the
logarithmic temperature scale). T$^{*}$ - characteristic
temperature associated with the "hump-like" feature in $\rho$(T),
possible origins of which are discussed in the text; T$_{1}$ -
magnetic ordering temperature; T$_{2}$ - ordering temperature
(probably magnetic). The vertical dashed line separates the T-P
phase diagram into two regions in which T$^{*}$, T$_{1}$, and
T$_{2}$ all have distinctly different pressure dependencies. We
grouped in a single low-temperature magnetic phase the critical
temperatures T$_{2}$ obtained from measurements of the electrical
resistivity under pressure and specific heat at ambient pressure,
although they might have different origins. $Inset$: room
temperature evolution of the electrical resistance as the pressure
was increased. The abrupt jump at 45 kbar may be due to a
structural phase transition.} \label{TvsP}
\end{figure}

A feature, occurring at a temperature denoted T$_{2,\rho}$, was
discovered above 70 kbar for the two crystals measured in the two
Bridgman experiments (Fig. \ref{RHOvsTvsP}, lower \textit{inset}).
The features are truncated at the base temperature of 1.1 K, where
the resistivity has dropped by 65\% of its value at the onset of the
transition. In the first Bridgman cell experiment, the drop of the
resistivity was detected at 2.7 K and 152 kbar and it was seen again
at 2.4 K and 74.5 kbar. For the second Bridgman run, T$_{2,\rho}$
remained at a value of 5.5 $\pm$ 0.1 K while increasing the pressure
from 86 to 124 kbar. This suggests a possible new phase emerging at
lower pressures and below the temperature range covered in this
experiment. Fig.~\ref{TvsP} summarizes the different regions of the
T-P phase diagram studied in the present work.

\begin{figure}[tbp]
{\includegraphics[width=3.4in]{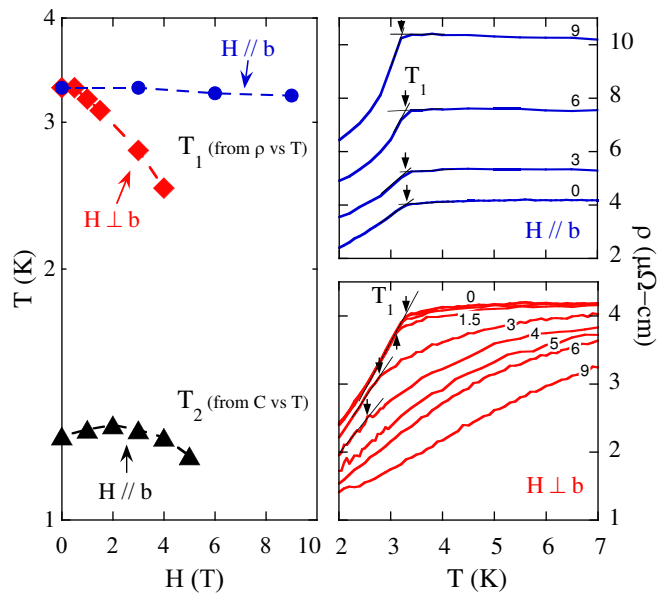}} \caption{(Color online)
\textit{Left}: Field dependence of the magnetic ordering
temperatures T$_{1,\rho}$ and T$_{2,C}$ of CeTe$_{3}$ for magnetic
fields applied perpendicular and parallel to the $b$-axis. A
logarithmic temperature scale was chosen to emphasize the
curvature of T$_{2}$ vs H. \textit{Right}: Temperature dependence
of the electrical resistivity near T$_{1}$ for magnetic fields
applied parallel (\textit{upper-right}; 0, 3, 6 and 9 T) and
perpendicular (\textit{lower-right}; 0, 0.5, 1, 1.5, 3, 4 ,5 ,6
and 9 T) to the \textit{b}-axis of the crystals. The arrows mark
the position of T$_{1}$. The numbers on each curve denote the
value of the applied magnetic field in tesla. For the
\textit{lower-right} panel, arrows and numbers for fields of 0.5
and 1 T are not displayed for better clarity.} \label{TvsH}
\end{figure}

Figure \ref{CvsT} displays the electronic and magnetic contributions
to the specific heat of CeTe$_{3}$ for magnetic fields up to 5 T
applied along the $b$-axis of the crystals, obtained after
subtracting the phonon contribution estimated from the high
temperature C(T) data. The C/T versus T$^{2}$ fits yielded a Debye
temperature of 161 K, comparable to previous values for LaTe$_{3}$,
and an electronic specific heat coefficient $\gamma$ of 52
mJ/mol-K$^{2}$, substantially larger than is observed for
LaTe$_{3}$,\cite{ru06} implying a moderately enhanced admixture of
the localized 4\textit{f} electron states of Ce with conduction
electron states, as suggested in previously reported angle-resolved
photoemission spectroscopy (ARPES) experiments.\cite{brouet08,
komoda04} A broad feature in C(T) characterizing the magnetic order
that occurs at T$_{1,C}$ corresponds directly to the transition
temperature T$_{1,\rho}$ obtained from electrical resistivity
measurements at ambient pressure ($\sim$ 3 K). This anomaly in the
specific heat remains unchanged by the magnetic fields used in this
experiment. The lower \textit{inset} in Fig. \ref{CvsT} shows the
electronic and magnetic contributions to the entropy at zero
magnetic field, which adds up to Rln2 (indicated by a horizontal
dashed line) at temperatures right above T$_{1,C}$, consistent with
what is expected for a Ce$^{3+}$-doublet ground state.

At even lower temperatures, a sharper feature is observed in the
C(T) data that exhibits a rather weak field dependence. This
transition was not detected in the electrical resistivity
experiments (down to 1.1 K). This suggests this new phase also has a
magnetic origin. The transition temperature T$_{2,C}$ (defined after
performing an equal-entropy analysis of the data) increases to a
value of $\sim$ 1.3 K at 2 T, and then decreases for the higher
applied fields. The left panel of Fig. \ref{TvsH} illustrates the
evolution of this feature throughout the range of applied magnetic
fields in which the specific heat measurements were made.

The above evidence associated with the low-temperature transition
below 1.5 K, and the apparent lack of field dependence for the 3 K
ordering temperature revealed by the specific heat data led us to
inquire further into the origin of these magnetic transitions. The
work by Iyeiri and co-workers previously mentioned, foretells a
strong dependence of the magnetic phases of CeTe$_{3}$ with the
orientation of the applied magnetic field with respect to the
crystalline axes. In order to test the angle dependence of T$_{1}$,
electrical resistivity measurements were performed down to 2 K in
magnetic fields $\leqslant$ 9 T applied perpendicular and parallel
to the $b$-axis of the crystals, and perpendicular to the direction
of the current passing through the the ac-planes of the samples,
utilizing a commercial Quantum Design PPMS sample rotator. The two
\textit{right} panels of Fig. \ref{TvsH} show these results. With
magnetic fields applied perpendicular to the planes (H $\parallel$
b, \textit{upper-right} panel), the transition temperature T$_{1}$
does not shift with applied magnetic field, consistent with the
specific heat measurements, although a rather strong
magnetoresistance was found (($R_{9T}-R_{0})/R_{0}=1.43$ at 10 K).
On the other hand, for fields applied parallel to the $ac$-planes
(\textit{lower-right} panel), a negative magnetoresistance is
observed, and the transition temperature T$_{1}$ moves towards zero
as indicated by the arrows. The \textit{left} panel in Fig.
\ref{TvsH} combines the field dependencies of T$_{1,\rho}$ for H
$\parallel$ $b$ and H $\perp$ $b$, with the field dependence of
T$_{2,C}$ with H $\parallel$ $b$.

Although not shown in Fig. \ref{TvsH} for clarity, the local
Kondo-like minimum around 10 K mentioned earlier\cite{ru06,
brouet08} has been seen in this set of measurements at ambient
pressure. For H $\parallel$ b, this minimum appears at 9.8 K without
applied magnetic field, and increases to 10.5, 12 and 13 K for 3, 6
and 9 T. For H $\perp$ $b$, the minimum observed at zero field at
9.8 K remains unchanged for fields below 1.5 T, and then disappears
for magnetic fields above 3 T.

The specific heat and transport data presented in Figs. \ref{CvsT}
and \ref{TvsH} suggest that T$_{1}$ characterizes the onset of the
transition to a magnetic phase with the \textit{easy} magnetic
axis contained in the $ac$-planes. The $b$-axis would then play
the role of a \textit{hard} axis for this magnetic phase,
consistent with magnetic susceptibility
measurements.\cite{iyeiri03} In the case of the transition at
T$_{2,C}$, we did not measure the specific heat with the magnetic
field applied parallel to the basal plane, but we can conclude
that the T$_{2,C}$ magnetic \textit{easy} axis is not parallel to
the T$_{1}$ easy axis, as can be found in other anisotropic
magnetic $f$-electron systems reported
elsewhere,\cite{elliott1961} in which ferromagnetic and
antiferromagnetic phases occur at different ordering temperatures,
with non-collinear ordering directions, due to the interaction of
the localized $f$-electrons with the conduction electrons
(intra-atomic exchange) and with nearby ions (interatomic
exchange), and to the effect of crystalline electric fields.

The temperature vs. pressure phase diagram is presented in Fig.
\ref{TvsP}. It is clearly seen that the curve  T$_{1}$(P) saturates
to a rather constant value above 50 kbar. T$^{*}$(P) also shows a
kink around the same pressure, and then attains a constant value of
$\sim$ 55 K. This particular value of pressure separates the phase
diagram in two regions: a low pressure region where the phase
characterized by T$^{*}$(P) competes with the phases below
T$_{1}$(P), and a high-pressure region where these three phases may
coexist. The \textit{inset} of Fig. \ref{TvsP} shows the
room-temperature dependence of the electrical resistance with
applied pressure (compression only). An abrupt drop of the
resistance occurs around 45 kbar, suggesting the existence of a
structural phase transition at this pressure. Despite the fact that
this feature in the electrical resistance was observed at room
temperature, the value of this pressure coincides with the kinks in
the T$^{*}$(P) and T$_{1}$(P)  curves at 50 kbar, and with the
emergence of the critical temperature T$_{2,\rho}$ above 60 kbar. As
we previously mentioned, in a recent work by Sacchetti \textit{et
al.},\cite{sacchetti08} x-ray diffraction experiments performed
under pressure showed that the satellite peaks associated with the
CDW lattice distortion disappear at room temperature when an
external pressure of 30 kbar is applied. This suggests that the
onset of the CDW transition is driven to lower temperatures when
high enough pressures are applied to the rare-earth tritellurides.
In the same report, the authors showed that the \textit{a} and
\textit{c} lattice parameters become equal at room temperature and
for pressures above 30 kbar which is indicative that a structural
phase transition might be taking place near that value of pressure.
Unfortunately, no x-ray data under pressure and at low temperatures
has yet been reported for CeTe$_{3}$, which would definitely clarify
the origin of T$^{*}$ that we found in our Bridgman cell
experiments.

\section{Conclusions}

In summary, we have presented high-pressure, transport and
thermodynamic measurements on CeTe$_{3}$ single crystals. These
measurements yielded evidence for two magnetic phases, detected in
electrical resistivity and specific heat measurements at low
temperatures, with non-parallel magnetic $easy$ axes. We also
reported the emergence of a phase at high pressures and low
temperatures and a possible structural transition detected at room
temperature and at 45 kbar, which could be related to the reduction
of the CDW transition temperature, illustrating that external
pressure plays a key role in establishing the phase diagram of the
highly anisotropic rare-earth tritellurides.

\section*{Acknowledgements}

Research at University of California, San Diego, was supported by
the National Nuclear Security Administration under the Stewardship
Science Academic Alliance Program through the U. S. Department of
Energy grant number DE-FG52-06NA26205. Crystal growth at Stanford
University was supported by the U. S. Department of Energy under
contract No. DE-AC02-76SF00515.


\begin{thebibliography}{30}
\expandafter\ifx\csname
natexlab\endcsname\relax\def\natexlab#1{#1}\fi
\expandafter\ifx\csname bibnamefont\endcsname\relax
   \def\bibnamefont#1{#1}\fi
\expandafter\ifx\csname bibfnamefont\endcsname\relax
   \def\bibfnamefont#1{#1}\fi
\expandafter\ifx\csname citenamefont\endcsname\relax
   \def\citenamefont#1{#1}\fi
\expandafter\ifx\csname url\endcsname\relax
   \def\url#1{\texttt{#1}}\fi
\expandafter\ifx\csname urlprefix\endcsname\relax\def\urlprefix{URL
}\fi \providecommand{\bibinfo}[2]{#2}
\providecommand{\eprint}[2][]{\url{#2}}

\bibitem{gruner94}
\bibinfo{author}{\bibfnamefont{G.}~\bibnamefont{Gr\"{u}ner}},
   \bibinfo{journal}\textit{Density Waves in Solids} (Addison-Wesley, Reading, MA, 1994).

\bibitem{dimasi95}
\bibinfo{author}{\bibfnamefont{E.}~\bibnamefont{DiMasi}},
   \bibinfo{author}{\bibfnamefont{M.}~\bibfnamefont{C.}~\bibnamefont{Aronson}},
   \bibinfo{author}{\bibfnamefont{J.}~\bibfnamefont{F.}~\bibnamefont{Mansfield}},
   \bibinfo{author}{\bibfnamefont{B.}~\bibnamefont{Foran}}, and
   \bibinfo{author}{\bibfnamefont{S.}~\bibnamefont{Lee}},
   \bibinfo{journal}{Phys. Rev. B} \textbf{\bibinfo{volume}{52}},
   \bibinfo{pages}{14516} (\bibinfo{year}{1995}).

\bibitem{brouet04}
\bibinfo{author}{\bibfnamefont{V.}~\bibnamefont{Brouet}},
   \bibinfo{author}{\bibfnamefont{W.}~\bibfnamefont{L.}~\bibnamefont{Yang}},
   \bibinfo{author}{\bibfnamefont{X.}~\bibfnamefont{J.}~\bibnamefont{Zhou}},
   \bibinfo{author}{\bibfnamefont{Z.}~\bibnamefont{Hussain}},
   \bibinfo{author}{\bibfnamefont{N.}~\bibnamefont{Ru}},
   \bibinfo{author}{\bibfnamefont{K.}~\bibfnamefont{Y.}~\bibnamefont{Shin}},
   \bibinfo{author}{\bibfnamefont{I.}~\bibfnamefont{R.}~\bibnamefont{Fisher}}, and
   \bibinfo{author}{\bibfnamefont{Z.}~\bibfnamefont{X.}~\bibnamefont{Shen}},
   \bibinfo{journal}{Phys. Rev. Lett.} \textbf{\bibinfo{volume}{93}},
   \bibinfo{pages}{126405} (\bibinfo{year}{2004}).

\bibitem{brouet08}
\bibinfo{author}{\bibfnamefont{V.}~\bibnamefont{Brouet}},
   \bibinfo{author}{\bibfnamefont{W.}~\bibfnamefont{L.}~\bibnamefont{Yang}},
   \bibinfo{author}{\bibfnamefont{X.}~\bibfnamefont{J.}~\bibnamefont{Zhou}},
   \bibinfo{author}{\bibfnamefont{Z.}~\bibnamefont{Hussain}},
   \bibinfo{author}{\bibfnamefont{R.}~\bibfnamefont{G.}~\bibnamefont{Moore}},
   \bibinfo{author}{\bibfnamefont{R.}~\bibnamefont{He}},
   \bibinfo{author}{\bibfnamefont{D.}~\bibfnamefont{H.}~\bibnamefont{Lu}},
   \bibinfo{author}{\bibfnamefont{Z.}~\bibfnamefont{X.}~\bibnamefont{Shen}},
   \bibinfo{author}{\bibfnamefont{J.}~\bibnamefont{Laverock}},
   \bibinfo{author}{\bibfnamefont{S.}~\bibfnamefont{B.}~\bibnamefont{Dugdale}},
   \bibinfo{author}{\bibfnamefont{N.}~\bibnamefont{Ru}}, and
   \bibinfo{author}{\bibfnamefont{I.}~\bibfnamefont{R.}~\bibnamefont{Fisher}},
   \bibinfo{journal}{Phys. Rev. B} \textbf{\bibinfo{volume}{77}},
   \bibinfo{pages}{235104} (\bibinfo{year}{2008}).

\bibitem{norling66}
\bibinfo{author}{\bibfnamefont{B.}~\bibfnamefont{K.}~\bibnamefont{Norling}},
   and
   \bibinfo{author}{\bibfnamefont{H.}~\bibnamefont{Steinfink}},
   \bibinfo{journal}{Inorg. Chem.} \textbf{\bibinfo{volume}{5}},
   \bibinfo{pages}{1488} (\bibinfo{year}{1966}).

\bibitem{komoda04}
\bibinfo{author}{\bibfnamefont{H.}~\bibnamefont{Komoda}},
   \bibinfo{author}{\bibfnamefont{T.}~\bibnamefont{Sato}},
   \bibinfo{author}{\bibfnamefont{S.}~\bibnamefont{Souma}},
   \bibinfo{author}{\bibfnamefont{T.}~\bibnamefont{Takahashi}},
   \bibinfo{author}{\bibfnamefont{Y.}~\bibnamefont{Ito}}, and
   \bibinfo{author}{\bibfnamefont{K.}~\bibnamefont{Suzuki}},
   \bibinfo{journal}{Phys. Rev. B} \textbf{\bibinfo{volume}{70}},
   \bibinfo{pages}{195101} (\bibinfo{year}{2004}).

\bibitem{pearsons97}
\bibinfo{author}{\bibnamefont{P.}~\bibnamefont{Villars}},
   \bibinfo{journal}\textit{Pearson's Handbook Desk Edition,
     Crystallographic Data for Intermetallic Phases} (ASM International, Materials Park, OH, 1997), Vol. 1, p. 1221.

\bibitem{sacchetti06}
\bibinfo{author}{\bibfnamefont{A.}~\bibnamefont{Sacchetti}},
   \bibinfo{author}{\bibfnamefont{L.}~\bibnamefont{Degiorgi}},
   \bibinfo{author}{\bibfnamefont{T.}~\bibnamefont{Giamarchi}},
   \bibinfo{author}{\bibfnamefont{N.}~\bibnamefont{Ru}}, and
   \bibinfo{author}{\bibfnamefont{I.}~\bibfnamefont{R.}~\bibnamefont{Fisher}},
   \bibinfo{journal}{Phys. Rev. B} \textbf{\bibinfo{volume}{74}},
   \bibinfo{pages}{125115} (\bibinfo{year}{2006}).

\bibitem{ru08}
\bibinfo{author}{\bibfnamefont{N.}~\bibnamefont{Ru}},
   \bibinfo{author}{\bibfnamefont{C.}~\bibfnamefont{L.}~\bibnamefont{Condron}},
   \bibinfo{author}{\bibfnamefont{G.}~\bibfnamefont{Y.}~\bibnamefont{Margulis}},
   \bibinfo{author}{\bibfnamefont{K.}~\bibfnamefont{Y.}~\bibnamefont{Shin}},
   \bibinfo{author}{\bibfnamefont{J.}~\bibnamefont{Laverock}},
   \bibinfo{author}{\bibfnamefont{S.}~\bibfnamefont{B.}~\bibnamefont{Dugdale}},
   \bibinfo{author}{\bibfnamefont{M.}~\bibfnamefont{F.}~\bibnamefont{Toney}}, and
   \bibinfo{author}{\bibfnamefont{I.}~\bibfnamefont{R.}~\bibnamefont{Fisher}},
   \bibinfo{journal}{Phys. Rev. B} \textbf{\bibinfo{volume}{77}},
   \bibinfo{pages}{035114} (\bibinfo{year}{2008}).

\bibitem{ru08_2}
\bibinfo{author}{\bibfnamefont{N.}~\bibnamefont{Ru}},
   \bibinfo{author}{\bibfnamefont{J.}~\bibfnamefont{-H.}~\bibnamefont{Chu}}, and
   \bibinfo{author}{\bibfnamefont{I.}~\bibfnamefont{R.}~\bibnamefont{Fisher}},
   \bibinfo{journal}{Phys. Rev. B} \textbf{\bibinfo{volume}{78}},
   \bibinfo{pages}{012410} (\bibinfo{year}{2008}).

\bibitem{sacchetti07}
\bibinfo{author}{\bibfnamefont{A.}~\bibnamefont{Sacchetti}},
   \bibinfo{author}{\bibfnamefont{E.}~\bibnamefont{Arcangeletti}},
   \bibinfo{author}{\bibfnamefont{A.}~\bibnamefont{Perucchi}},
   \bibinfo{author}{\bibfnamefont{L.}~\bibnamefont{Baldassarre}},
   \bibinfo{author}{\bibfnamefont{P.}~\bibnamefont{Postorino}},
   \bibinfo{author}{\bibfnamefont{S.}~\bibnamefont{Lupi}},
   \bibinfo{author}{\bibfnamefont{N.}~\bibnamefont{Ru}},
   \bibinfo{author}{\bibfnamefont{I.}~\bibfnamefont{R.}~\bibnamefont{Fisher}}, and
   \bibinfo{author}{\bibfnamefont{L.}~\bibnamefont{Degiorgi}}
   \bibinfo{journal}{Phys. Rev. Lett.} \textbf{\bibinfo{volume}{98}},
   \bibinfo{pages}{026401} (\bibinfo{year}{2007}).

\bibitem{lavagnini09}
\bibinfo{author}{\bibfnamefont{M.}~\bibnamefont{Lavagnini}},
   \bibinfo{author}{\bibfnamefont{A.}~\bibnamefont{Sacchetti}},
   \bibinfo{author}{\bibfnamefont{C.}~\bibnamefont{Marini}},
   \bibinfo{author}{\bibfnamefont{M.}~\bibnamefont{Valentini}},
   \bibinfo{author}{\bibfnamefont{R.}~\bibnamefont{Sopracase}},
   \bibinfo{author}{\bibfnamefont{A.}~\bibnamefont{Perucchi}},
   \bibinfo{author}{\bibfnamefont{P.}~\bibnamefont{Postorino}},
   \bibinfo{author}{\bibfnamefont{S.}~\bibnamefont{Lupi}},
   \bibinfo{author}{\bibfnamefont{J.}~\bibfnamefont{H.}~\bibnamefont{Chu}},
   \bibinfo{author}{\bibfnamefont{I.}~\bibfnamefont{R.}~\bibnamefont{Fisher}}, and
   \bibinfo{author}{\bibfnamefont{L.}~\bibnamefont{Degiorgi}}
   \bibinfo{journal}{Phys. Rev. B.} \textbf{\bibinfo{volume}{79}},
   \bibinfo{pages}{075117} (\bibinfo{year}{2009}).

\bibitem{sullow99}
\bibinfo{author}{\bibfnamefont{S.}~\bibnamefont{S\"{u}llow}},
   \bibinfo{author}{\bibfnamefont{M.}~\bibfnamefont{C.}~\bibnamefont{Aronson}},
   \bibinfo{author}{\bibfnamefont{B.}~\bibfnamefont{D.}~\bibnamefont{Rainford}}, and
   \bibinfo{author}{\bibfnamefont{P.}~\bibnamefont{Haen}},
   \bibinfo{journal}{Phys. Rev. Lett.} \textbf{\bibinfo{volume}{82}},
   \bibinfo{pages}{2963} (\bibinfo{year}{1999}).

\bibitem{jung03}
\bibinfo{author}{\bibfnamefont{M.}~\bibfnamefont{H.}~\bibnamefont{Jung}},
   \bibinfo{author}{\bibfnamefont{A.}~\bibnamefont{Alsmadi}},
   \bibinfo{author}{\bibfnamefont{H.}~\bibfnamefont{C.}~\bibnamefont{Kim}},
   \bibinfo{author}{\bibfnamefont{Yunkyu}~\bibnamefont{Bang}},
   \bibinfo{author}{\bibfnamefont{K.}~\bibfnamefont{H.}~\bibnamefont{Ahn}},
   \bibinfo{author}{\bibfnamefont{K.}~\bibnamefont{Umeo}},
   \bibinfo{author}{\bibfnamefont{A.}~\bibfnamefont{H.}~\bibnamefont{Lacerda}},
   \bibinfo{author}{\bibfnamefont{H.}~\bibnamefont{Nakotte}},
   \bibinfo{author}{\bibfnamefont{H.}~\bibfnamefont{C.}~\bibnamefont{Ri}},
   and
   \bibinfo{author}{\bibfnamefont{T.}~\bibnamefont{Takabatake}},
   \bibinfo{journal}{Phys. Rev. B} \textbf{\bibinfo{volume}{67}},
   \bibinfo{pages}{212504} (\bibinfo{year}{2003}).

\bibitem{mathur98}
\bibinfo{author}{\bibfnamefont{N.}~\bibfnamefont{D.}~\bibnamefont{Mathur}},
   \bibinfo{author}{\bibfnamefont{F.}~\bibfnamefont{M.}~\bibnamefont{Grosche}},
   \bibinfo{author}{\bibfnamefont{S.}~\bibfnamefont{R.}~\bibnamefont{Julian}},
   \bibinfo{author}{\bibfnamefont{I.}~\bibfnamefont{R.}~\bibnamefont{Walker}},
   \bibinfo{author}{\bibfnamefont{D.}~\bibfnamefont{M.}~\bibnamefont{Freye}},
   \bibinfo{author}{\bibfnamefont{R.}~\bibfnamefont{K.}~\bibfnamefont{W.}~\bibnamefont{Haselwimmer}},
   and
   \bibinfo{author}{\bibfnamefont{G.}~\bibfnamefont{G.}~\bibnamefont{Lonzarich}},
   \bibinfo{journal}{Nature} \textbf{\bibinfo{volume}{394}},
   \bibinfo{pages}{39} (\bibinfo{year}{1998}).

\bibitem{berthier76}
\bibinfo{author}{\bibfnamefont{C.}~\bibnamefont{Berthier}},
   \bibinfo{author}{\bibfnamefont{P.}~\bibnamefont{Molini\'e}}, and
   \bibinfo{author}{\bibfnamefont{D.}~\bibnamefont{G\'erome}},
   \bibinfo{journal}{Solid State Commun.} \textbf{\bibinfo{volume}{18}},
   \bibinfo{pages}{1393} (\bibinfo{year}{1976}).

\bibitem{ru06}
\bibinfo{author}{\bibfnamefont{N.}~\bibnamefont{Ru}} and
   \bibinfo{author}{\bibfnamefont{I.}~\bibfnamefont{R.}~\bibnamefont{Fisher}},
   \bibinfo{journal}{Phys. Rev. B} \textbf{\bibinfo{volume}{73}},
   \bibinfo{pages}{033101} (\bibinfo{year}{2006}).

\bibitem{bireckoven88}
\bibinfo{author}{\bibfnamefont{B.}~\bibnamefont{Bireckoven}} and
   \bibinfo{author}{\bibfnamefont{J.}~\bibnamefont{Wittig}},
   \bibinfo{journal}{J. Phys. E: Sci. Instrum.} \textbf{\bibinfo{volume}{21}},
   \bibinfo{pages}{841} (\bibinfo{year}{1988}).

\bibitem{sacchetti08}
\bibinfo{author}{\bibfnamefont{A.}~\bibnamefont{Sacchetti}},
   \bibinfo{author}{\bibfnamefont{C.}~\bibfnamefont{L.}~\bibnamefont{Condron}},
   \bibinfo{author}{\bibfnamefont{S.}~\bibfnamefont{N.}~\bibnamefont{Gvasaliya}},
   \bibinfo{author}{\bibfnamefont{F.}~\bibnamefont{Pfuner}},
   \bibinfo{author}{\bibfnamefont{M.}~\bibnamefont{Lavagnini}},
   \bibinfo{author}{\bibfnamefont{M.}~\bibnamefont{Baldini}},
   \bibinfo{author}{\bibfnamefont{M.}~\bibfnamefont{F.}~\bibnamefont{Toney}},
   \bibinfo{author}{\bibfnamefont{M.}~\bibnamefont{Merlini}},
   \bibinfo{author}{\bibfnamefont{M.}~\bibnamefont{Hanfland}},
   \bibinfo{author}{\bibfnamefont{J.}~\bibnamefont{Mesot}},
   \bibinfo{author}{\bibfnamefont{J.}~\bibfnamefont{-H.}~\bibnamefont{Chu}}
   \bibinfo{author}{\bibfnamefont{I.}~\bibfnamefont{R.}~\bibnamefont{Fisher}},
   \bibinfo{author}{\bibfnamefont{P.}~\bibnamefont{Postorino}}, and
   \bibinfo{author}{\bibfnamefont{L.}~\bibnamefont{Degiorgi}},
   \bibinfo{journal}{arXiv:0811.0338v1} (\bibinfo{year}{unpublished}).

\bibitem{iyeiri03}
\bibinfo{author}{\bibfnamefont{Y.}~\bibnamefont{Iyeiri}},
   \bibinfo{author}{\bibfnamefont{T.}~\bibnamefont{Okumura}},
   \bibinfo{author}{\bibfnamefont{C.}~\bibnamefont{Michioka}}, and
   \bibinfo{author}{\bibfnamefont{K.}~\bibnamefont{Suzuki}},
   \bibinfo{journal}{Phys. Rev. B} \textbf{\bibinfo{volume}{67}},
   \bibinfo{pages}{144417} (\bibinfo{year}{2003}).

\bibitem{elliott1961}
\bibinfo{author}{\bibfnamefont{R.}~\bibfnamefont{J.}~\bibnamefont{Elliott}},
   \bibinfo{journal}{Phys. Rev.} \textbf{\bibinfo{volume}{124}},
   \bibinfo{pages}{346} (\bibinfo{year}{1961}).



\end{thebibliography}
\end{document}